\let\latexaddtocontents\addtocontents
\documentclass[floatfix,aps,prl,twocolumn,10pt,superscriptaddress]{revtex4-2}
\let\addtocontents\latexaddtocontents

\usepackage{amsmath}
\usepackage{graphicx} 
\usepackage{subfigure}
\usepackage{xcolor}
\usepackage{centernot}
\usepackage{mathtools}
\usepackage[normalem]{ulem}
\begin{document}

\def\red{\textcolor{red}}
\def\blue{\textcolor{blue}}

\title{Added mass effect in coupled Brownian particles}
\author{Long Him Cheung }
\email{lhcheung@umd.edu}
\affiliation{Department of Physics, University of Maryland, College Park, MD 20742 USA}
\author{Christopher Jarzynski}
\email{cjarzyns@umd.edu}
\affiliation{Department of Physics, University of Maryland, College Park, MD 20742 USA}
\affiliation{Department of Chemistry and Biochemistry, University of Maryland, College Park, MD 20742 USA}
\affiliation{
Institute for Physical Science and Technology, University of Maryland, College Park, MD 20742, USA}
\date{\today}

\begin{abstract}
The {\it added mass effect} is the contribution to a Brownian particle's effective mass arising from the hydrodynamic flow its motion induces.
For a spherical particle in an incompressible fluid, the added mass is half the fluid's displaced mass, but in a compressible fluid its value depends on a competition between timescales.
Here we illustrate this behavior with a solvable model of two harmonically coupled Brownian particles of mass $m$, one representing the sphere, the other the immediately surrounding fluid.
The measured distribution of the Brownian particle's velocity, $P(\bar{v})$, follows a Maxwell-Boltzmann distribution with an effective mass $m^*$.
Solving analytically for $m^*$, we find that its value is determined by three relevant timescales: the momentum relaxation time, $t_p$, the harmonic oscillation period, $\tau$, and the velocity measurement time resolution, $\Delta t$.
In limiting cases $\Delta t \ll \tau,t_p$ and $\tau\ll\Delta t\ll t_p$, our expression for $m^*$ reduces to $m$ and $2m$, respectively.
We find similar behavior upon generalizing the model to the case of unequal masses.
\end{abstract}

\maketitle

\section{Introduction}
Brownian motion, that is the random movement of a particle suspended in a liquid or gas, was argued theoretically by Sutherland~\cite{Sutherland1905}, Einstein~\cite{Einstein1905} and Smoluchowski~\cite{Smoluchowski1906}, and confirmed experimentally by Perrin~\cite{Perrin1909}, to arise from the particle's collisions with the surrounding fluid's molecules.
In equilibrium at temperature $T$, the $D$-dimensional velocity ${\bf v}$ of a Brownian particle with mass $m$ obeys the Maxwell-Boltzmann distribution,
\begin{equation}
\label{eq:MBD}
   P^\textrm{MB}({\bf v}) =\left(\frac{m\beta}{2\pi}\right)^{D/2} e^{-\beta m {\bf v^2} / 2}
    \quad,\quad \beta = \frac{1}{k_BT}
\end{equation}
which in turn implies the equipartition theorem: the average kinetic energy per degree of freedom is $1/2\beta$. Because the instantaneous velocity ${\bf v}(t)$ randomizes quickly, the direct measurement of ${\bf v}(t)$ requires fine temporal and spatial resolutions.
These experimental challenges have been overcome only recently, by Li, Mo, Raizen and colleagues, first for a Brownian particle immersed in gas~\cite{Li2010}, then in liquid~\cite{Mo2015}, marking milestones in the precision testing of fundamental statistical mechanics.

For Brownian motion in liquid surroundings, the velocity ${\bf v}$ was observed to obey a modified Maxwell-Boltzmann distribution~\cite{Mo2015}, with the particle's mass $m$ in Eq.~\ref{eq:MBD} replaced by an effective mass
\begin{equation}
\label{eq:mstar}
    m^* \approx
    m + \frac{1}{2} M_d \quad ,
\end{equation}
where $M_d$ is the mass of liquid displaced by the particle.
While this result may seem to conflict with classical statistical mechanics, the discrepancy is understood to arise from hydrodynamic considerations~\cite{Zwanzig1975,Brennen1982}. As the particle moves with speed $v$, the surrounding fluid flows around it. If the particle is spherical and the fluid incompressible, then the induced flow has a kinetic energy $(1/4)M_d v^2$, giving rise to the added mass $M_d/2$ in Eq.~\ref{eq:mstar}.

At finite fluid compressibility, the effective mass is determined by a competition between two timescales: a characteristic time $\tau_{\textrm{fluid}}\sim R/c$ for the fluid to respond to displacements of the Brownian particle (where $R$ is the particle's radius and $c$ the speed of sound), and the time resolution $\Delta t$ with which the time-averaged velocity $\bar{\bf v}=\Delta{\bf q}/\Delta t$ is measured \cite{Zwanzig1975,Mo2019}. If the velocity is measured with arbitrarily precise time resolution, such that $\Delta t \ll \tau_{\textrm{fluid}}$, then the effective mass is the particle's true mass, $m^*=m$, thus recovering the ordinary Maxwell-Boltzmann distribution; while if $\tau_{\textrm{fluid}} \ll \Delta t$, the effective mass is given by the right side of Eq.~\ref{eq:mstar}. In this paper, we analyze an exactly solvable model to illustrate this behavior, and to quantitatively describe the crossover between these two regimes.

Our model consists of two Brownian particles of equal mass $m$ moving in one dimension, coupled through a harmonic spring and interacting with a thermal environment.
One of these particles plays the role of the Brownian particle described in the previous paragraphs. The other represents, roughly, the immediately surrounding fluid.
The spring is analogous to the coupling between the Brownian particle and the fluid.
We imagine that the first particle's position is measured at regularly spaced times, with $\Delta t$ the interval between successive measurements, and $\Delta q_1$ the displacement over one such interval. The time-averaged velocity $\bar v_1 = \Delta q_1/\Delta t$ then represents a single measurement of velocity. An empirical velocity distribution $P(\bar{v}_1)$ is constructed from many such successive measurements.

A spring constant $k$ quantifies the harmonic coupling strength.
If the coupling is loose ($k\approx 0$) then the particles' motions are not strongly correlated, and we intuitively expect velocity measurements performed on the first particle to produce a Maxwell-Boltzmann distribution with effective mass $m$.
In the opposite extreme of stiff coupling ($k\rightarrow\infty$), the particles become ``glued together'' and we expect to observe a Maxwell-Boltzmann distribution with effective mass $2m$.
The particles' synchronized fluctuations in the stiff-coupling limit are analogous to the instantaneous flow induced by a Brownian particle in an incompressible fluid.

Our analysis will show that the empirical distribution $P(\bar{v}_1)$ is indeed a modified Maxwell-Boltzmann distribution, with an effective mass $m^*$ that depends on the interplay between three timescales: the momentum relaxation time $t_p=m/\gamma$, where $\gamma$ is a friction coefficient; the harmonic oscillation period $\tau=2\pi\sqrt{m/2k}$, analogous to the fluid response time $\tau_{\textrm{fluid}}$ discussed above; and the measurement time interval $\Delta t$.
We assume the velocity is measured faster than it randomizes, i.e.\ $\Delta t\ll t_p$, corresponding to the experimental conditions of Refs.~\cite{Li2010,Mo2015}.
We then find that when $\tau \ll \Delta t \ll t_p$ the effective mass is $m^*\approx2m$, whereas when $\Delta t \ll \tau  \ll t_p$ or $\Delta t \ll t_p \ll \tau$ we obtain $m^*\approx m$.

\section{Model and Analysis}
\label{sec:model}

Consider two identical, underdamped Brownian particles of mass $m$, moving in one dimension, immersed in a thermal medium with friction coefficient $\gamma$ and inverse temperature $\beta$, and connected by a spring of stiffness $k$. The equations of motion are:
\begin{subequations}
\label{eq:Eq_of_Motion_Same}
\begin{align}
    m\ddot{q}_1 &= -k(q_1-q_2) - \gamma \dot{q}_1 + \sqrt{2\gamma/\beta}\,\xi_1 \\
    m\ddot{q}_2 &= -k(q_2-q_1) - \gamma \dot{q}_2 + \sqrt{2\gamma/\beta}\,\xi_2 \quad ,
\end{align}
\end{subequations}
where $q_1$ and $q_2$ are the particles' positions, $\xi_1$ and $\xi_2$ are independent realizations of delta-correlated Gaussian white noise with zero mean and unit variance,
\begin{equation}
\label{eq:Gaussian_White_Noise}
     \langle \xi_{i}(t) \rangle = 0\, , \, \langle \xi_{i}(t)\,\xi_{j}(s) \rangle = \,\delta_{ij}\delta (t-s) \quad ,
\end{equation}
and the magnitude of the noise $\sqrt{2\gamma/\beta}$ follows from the fluctuation-dissipation theorem.
Under these dynamics, the distribution of each particle's velocity, $v_i = \dot{q}_i$, relaxes to the Maxwell-Boltzmann distribution corresponding to the true particle mass $m$:
\begin{equation}
\label{eq:MBD_i}
    P^\textrm{MB}(v_i) \propto e^{-\beta m v^2_i/2} \quad ,
\end{equation}
whose variance is $\sigma_{v_i}^2 = 1/\beta m$.

In an experiment, one does not directly measure a particle's velocity but rather its displacement $\Delta q_i$ over a time interval $\Delta t$. The time-averaged velocity
\begin{equation}
\label{eq:avg_velocity}
    \bar{v}_i = \frac{\Delta q_i}{\Delta t}
\end{equation}
converges to the instantaneous velocity when $\Delta t \to 0$, but in practice $\Delta t$ remains finite due to the limited time resolution of the measurement device. As a result, the empirically measured distribution $P(\bar{v}_i)$ differs from $P^\textrm{MB}(v_i)$ if $\Delta t$ is not sufficiently small to resolve all relevant velocity fluctuations.

If the measured velocity distribution $P(\bar{v}_i)$ is a Gaussian with zero mean (as we shall show to be the case) and variance $\sigma_{\bar{v}_i}^2$, then it can be viewed as a modified Maxwell-Boltzmann distribution with an effective mass
\begin{equation}
\label{eq:mstar-in-terms-of-variance}
    m^*=\frac{1}{\beta\sigma_{\bar{v}_i}^2} \quad .
\end{equation}
Our aim is to solve for $P(\bar{v}_i)$ for our simple model, and to explore how the resulting effective mass $m^*$ depends on the parameters $m,\gamma,k$, and (especially) $\Delta t$. We will imagine that the experimentalist tracks the position of particle 1 only and not particle 2, with a regular measurement time interval $\Delta t$. Hence we will focus on $P_{\Delta t}(\bar{v}_1)$, where the notation emphasizes that the empirically measured velocity distribution of particle 1 depends on $\Delta t$.

Since $\Delta t$ is fixed, a change of variables gives
\begin{equation}
\label{eq:Change_of_Var_Vbar_to_Deltaq}
    P_{\Delta t}(\bar{v}_1) = P_{\Delta t}(\Delta q_1)\Delta t \quad ,
\end{equation}
where $P_{\Delta t}(\Delta q_1)$ is the measured distribution of displacements $\Delta q_1 = \bar v_1 \Delta t$. Next, define $P_t(q_1|q_{10})$ to be the conditional probability to find particle 1 at $q_1$ at time $t$, given an initial position $q_{10}$ at time 0, i.e. $P_{t=0}(q_1|q_{10})=\delta(q_1-q_{10})$. As we will show, if the two-particle system is in equilibrium, then
\begin{equation}
\label{eq:P(q_1|q_10)=P(Delta_q1)}
    P_{t=\Delta t}(q_1|q_{10}) = P_{\Delta t}(\Delta q_1) \quad , 
\end{equation}
with $\Delta q_1=q_1-q_{10}$.
In other words, the distribution of displacements is independent of the particle's initial location. Hence, assuming the system has equilibrated, the problem of computing $P_{\Delta t}(\bar{v}_1)$ reduces to that of solving for $P_t(q_1|q_{10})$.

Introducing the center of mass $Q = (q_1+q_2)/2$, separation $q = q_1 - q_2$, and corresponding velocities $V$ and $v$, Eq.~\ref{eq:Eq_of_Motion_Same} can be rewritten as four first-order equations:
\begin{subequations}
\label{eq:Decoupled_Eq_of_Motion}
\begin{align}
    \dot Q &= V \label{eq:Decoupled_Eq_of_Motion_for_Q} \\
    \dot{V} & = -\frac{\gamma}{m} V + \frac{\sqrt{2\gamma/\beta}}{2m}(\xi_1+\xi_2) \label{eq:Decoupled_Eq_of_Motion_for_V} \\
    \dot q &= v \label{eq:Decoupled_Eq_of_Motion_for_q} \\
    \dot{v} & = -\omega^2 q -\frac{\gamma}{m} v + \frac{\sqrt{2\gamma/\beta}}{m}(\xi_1-\xi_2) \label{eq:Decoupled_Eq_of_Motion_for_v}
\end{align}
\end{subequations}
with
\begin{equation}
\label{eq:omegadef}
    \omega^2 = \frac{2k}{m} \quad .
\end{equation}
Since these dynamics are linear in $Q$, $V$, $q$ and $v$, with added Gaussian white noise, and since Eqs.~\ref{eq:Decoupled_Eq_of_Motion_for_Q} and \ref{eq:Decoupled_Eq_of_Motion_for_V} are decoupled from Eqs.~\ref{eq:Decoupled_Eq_of_Motion_for_q} and \ref{eq:Decoupled_Eq_of_Motion_for_v}, the conditional joint probability distributions $P_t(Q,V|Q_0,V_0)$ and $P_t(q,v|q_0,v_0)$ are both bivariate Gaussians:
\begin{subequations}
\label{eq:P(Q,V|Q_0,V_0)&P(q,v|q_0,v_0)=Gaussian}
    \begin{align}
    P_t(Q,V|Q_0,V_0) & =  \frac{1}{2\pi\sqrt{|\mathbf{C}|}}\exp\Big(-\frac{1}{2}\mathbf{X^{T} C^{-1} X}\Big) \\
    P_t(q,v|q_0,v_0) & =  \frac{1}{2\pi\sqrt{|\mathbf{c}|}}\exp\Big(-\frac{1}{2}\mathbf{x^{T}c^{-1}x}\Big) 
    \end{align}
\end{subequations}
with
\begin{equation} 
    \mathbf{X} = \begin{pmatrix} Q - \langle Q \rangle \\ V - \langle V \rangle \end{pmatrix} \quad,\quad  \mathbf{x} = \begin{pmatrix} q - \langle q \rangle \\ v - \langle v \rangle \end{pmatrix} \quad .
\end{equation}
Here $Q_0$, $V_0$, $q_0$ and $v_0$ denote initial positions and velocities, angular brackets $\langle \, \cdot \, \rangle$ denotes an ensemble average, and $\mathbf{C}(t)$ and $\mathbf{c}(t)$ are the covariance matrices for $(Q,V)$ and $(q,v)$ respectively. Explicit expressions for $\langle Q\rangle$, $\langle V\rangle$, $\langle q\rangle$, $\langle v\rangle$, $\mathbf{C}$, and $\mathbf{c}$ are given by Eqs.~\ref{eq:Mean_for_Q} -\ref{eq:covariance-matrix-of-qv-appendix} in the Appendix.

Now assume that the particles' velocities have equilibrated prior to $t=0$, hence $V_0$ and $v_0$ are sampled from equilibrium.
We then integrate over all velocity variables in Eq.~\ref{eq:P(Q,V|Q_0,V_0)&P(q,v|q_0,v_0)=Gaussian} to obtain the conditional distributions
\begin{subequations}
\label{eq:P(Q|Q_0)&P(q|q_0)-text}
    \begin{align}
    P_t(Q|Q_0) & = \frac{1}{\sqrt{2\pi\sigma^2_{Q}} }\exp\Big(-\frac{1}{2\sigma^2_Q}\big(Q - \bar{Q}\big)^2\Big) \\
    P_t(q|q_0) & =  \frac{1}{\sqrt{2\pi\sigma^2_{q}} }\exp\Big(-\frac{1}{2\sigma^2_q}\big(q - \bar{q}\big)^2\Big)
\end{align}
\end{subequations}
(see Appendix for details) with
\begin{align}
\label{eq:Qbar-and-qbar}
\bar{Q} &= Q_0 \,\, , \,\, \sigma^2_{Q} = \frac{1}{\beta \gamma} \Big( t - \frac{m}{\gamma}+\frac{m}{\gamma}e^{-\gamma t/m} \Big)\ \nonumber\\
\bar{q} &= \alpha q_0 \,\, , \,\, \sigma^2_{q} = \frac{2}{\beta m \omega^2}\big(1-\alpha^2\big) 
\nonumber \\
\alpha  &= \frac{\lambda_+ e^{-\lambda_-t}-\lambda_-e^{-\lambda_+t}}{\lambda_+ - \lambda_-} \nonumber\\
\lambda_\pm  & = \frac{1}{2}\Bigl(\frac{\gamma}{m} \pm \sqrt{\frac{\gamma^2}{m^2}-4\omega^2}\Bigr) \quad .
\end{align}
Since $P_t(Q|Q_0)$ and $P_t(q|q_0)$ are Gaussians, and since $q_1 = Q + q/2$ is the sum of the statistically independent random variables $Q$ and $q/2$,  it follows that $P_t(q_1|Q_0,q_0)$ is also a Gaussian,
\begin{equation}
\label{eq:P(q1|Q_0,q_0)}
    P_t(q_1|Q_0,q_0) = \frac{1}{\sqrt{2\pi\sigma^2_{q_1}}} \exp\Big( -\frac{1}{2\sigma^2_{q_1}}\big(q_1 - \bar{q}_1 \big)^2 \Big) 
\end{equation}
with mean $\bar{q}_1 = \bar{Q}+\bar{q}/2$ and variance $\sigma^2_{q_1} = \sigma^2_{Q}+ \sigma^2_{q}/4$.
Eq.~\ref{eq:P(q1|Q_0,q_0)} gives the probability distribution to find particle 1 at location $q_1$ at time $t$, conditioned on the initial values of the center of mass and separation at time $0$.

Note from Eqs.~\ref{eq:P(Q|Q_0)&P(q|q_0)-text} and \ref{eq:Qbar-and-qbar} that in the long-time limit, the center of mass $Q$ evolves diffusively ($\sigma_Q^2 \propto t$) whereas the separation $q$ settles to an equilibrium distribution with zero mean and variance $2/\beta m\omega^2$.
Let us assume that this equilibration occurs prior to $t=0$ (as we did earlier with the velocities), so that the separation $q_0$ is sampled from equilibrium.
Furthermore, let us perform a change of variables from $P_t(q_1|Q_0,q_0)$ to $P_t(q_1|q_{10},q_0)$, where $q_{10} = Q_0 + q_0/2$ is the initial value of $q_1$, and let us integrate over $q_0$ (sampled from equilibrium) to obtain $P_t(q_{1}|q_{10})$. Again leaving the details to the Appendix, we state the result:
\begin{equation}
\label{eq:P_t(q1|q_10)}
    P_t(q_1|q_{10}) = \frac{1}{\sqrt{2\pi \sigma_{\Delta q_1}^2(t)}}\exp\Big(-\frac{1}{2\sigma_{\Delta q_1}^2(t)}\big(q_1-q_{10}\big)^2\Big) 
\end{equation}
with
\begin{subequations}
\label{eq:sigma_Exact_Same}
\begin{align}
        \sigma_{\Delta q_1}^2(t) &= \frac{t}{\beta \gamma}+\frac{1}{2\beta m \omega^2} \Big\{ 2 - \frac{2m^2\omega^2}{\gamma^2}\big( 1-e^{-\gamma t/m} \big) \nonumber \\
        &- 2e^{-\gamma t/2m}\Big[\frac{\gamma}{ma}\sinh\left(\frac{at}{2}\right)+\cosh\left(\frac{at}{2}\right)\Big]\Big\} \\
     a &= \sqrt{\frac{\gamma^2}{m^2}-4\omega^2} \,\, = \,\, \sqrt{\frac{\gamma^2}{m^2}-\frac{8k}{m}} \quad .
\end{align}
\end{subequations}

We now return to the scenario in which the experimentalist tracks particle 1 by measuring its location at regular time intervals $\Delta t$.
Eq.~\ref{eq:P_t(q1|q_10)} shows that the particle's displacement during one interval, $\Delta q_1 = q_1 - q_{10}$, is statistically independent of its initial location $q_{10}$, reflecting the problem's underlying translational symmetry.
It follows that the displacements $\Delta q_1$ during successive time intervals are independent samples from the distribution
\begin{equation}
\label{eq:P_Delta t(Delta q_1)}
    P_{\Delta t}(\Delta q_1) = \frac{1}{\sqrt{2\pi \sigma_{\Delta q_1}^2(\Delta t)}}\exp\Big(-\frac{1}{2\sigma_{\Delta q_1}^2(\Delta t)}\Delta q_1^2\Big)
\end{equation}
with $\sigma_{\Delta q_1}^2(\Delta t)$ given by Eq.~\ref{eq:sigma_Exact_Same}.

Eqs.~\ref{eq:Change_of_Var_Vbar_to_Deltaq} and \ref{eq:P_Delta t(Delta q_1)} show that the empirically measured distribution of particle 1's velocity, $P_{\Delta t}(\bar v_1)$, is a Gaussian with zero mean and variance $\sigma_{\bar v_1}^2 = \sigma_{\Delta q_1}^2(\Delta t)/\Delta t^2$.
As already mentioned this distribution can be interpreted as a modified Maxwell-Boltzmann distribution with an effective mass $m^*=1/\beta\sigma_{\bar v_1}^2 = \Delta t^2/\beta \sigma_{\Delta q_1}^2(\Delta t)$ (see Eq.~\ref{eq:mstar-in-terms-of-variance}).
We thus finally arrive at our main result:
\begin{subequations}
\label{eq:mstar_Exact_Same}
    \begin{align}
    m^* &= \bigg( \frac{ t_p}{ m \Delta t}+\frac{\tau^2}{4\pi^2m\Delta t^2} \Big\{ 1 - \frac{4\pi^2 t_p^2}{\tau^2} \big( 1-e^{-\Delta t/t_p} \big) \nonumber\\
        &- e^{-\Delta t/2t_p}\Big[\frac{1}{at_p}\sinh\big(\frac{a\Delta t}{2}\big)+\cosh\big(\frac{a\Delta t}{2}\big)\Big]\Big\} \bigg)^{-1} \\
 a &= \sqrt{\frac{1}{t_p^2} -\frac{16\pi^2}{\tau^2} } \quad, \quad t_p = \frac{m}{\gamma} \quad,\quad \tau = \frac{2\pi}{\omega}  \quad ,
\end{align}
\end{subequations}
which gives the effective mass $m^*$ in terms of the true mass $m$, and three timescales: the measurement time $\Delta t$, the momentum relaxation time $t_p$, and the oscillation period $\tau$.

Eq.~\ref{eq:mstar_Exact_Same} is exact but complicated.
It simplifies greatly if we assume the timescales $\Delta t$, $t_p$ and $\tau$ are widely separated.
For $\bar{v}_1$ to provide a reasonable estimate of the instantaneous velocity $v_1$, a minimal requirement is that $\Delta t \ll t_p$: repeated measurements of position must be made before thermal noise randomizes the particle's momentum.
Under this assumption, as shown in the Appendix, the value of $m^*$ is approximately either $2m$ or $m$, depending on the interplay between $\Delta t$ and $\tau$.
Specifically, we identify three regimes:
\begin{subequations}
 \label{eq:mstar_Regime_Same}   
\begin{align}
    \label{eq:regime1} \textrm{regime 1} &: \tau \ll \Delta t \ll t_p \quad \to \quad  m^*  \approx 2m \\
    \label{eq:regime2} \textrm{regime 2}&: \Delta t \ll \tau \ll t_p \quad \to \quad m^* \approx m  \\
    \label{eq:regime3} \textrm{regime 3} &: \Delta t \ll t_p \ll \tau \quad \to \quad m^* \approx m \quad .
\end{align}
\end{subequations}

These results can be understood intuitively.
Regime 1, in which the oscillation period $\tau$ is the shortest timescale, represents the limit of large spring stiffness, $k\rightarrow\infty$.
In this limit the two Brownian particles are effectively stuck together and move as one object of mass $2m$.
Although particle 1 oscillates rapidly (as does particle 2), these oscillations are not resolved by measurements occurring at intervals $\Delta t$.
In regimes 2 and 3, $\Delta t$ is the shortest timescale, hence measurements of particle 1's position are able to resolve its instantaneous velocity.
The difference between regimes 2 and 3 is that the former ($\tau\ll t_p$) represents underdamped motion -- the particle separation $q$ exhibits recognizable oscillations -- while the latter ($t_p\ll \tau$) corresponds to overdamped motion, in which each particle's momentum thermally randomizes before oscillations occur.

\begin{figure}[!hbtp]
    \includegraphics[width=\columnwidth]{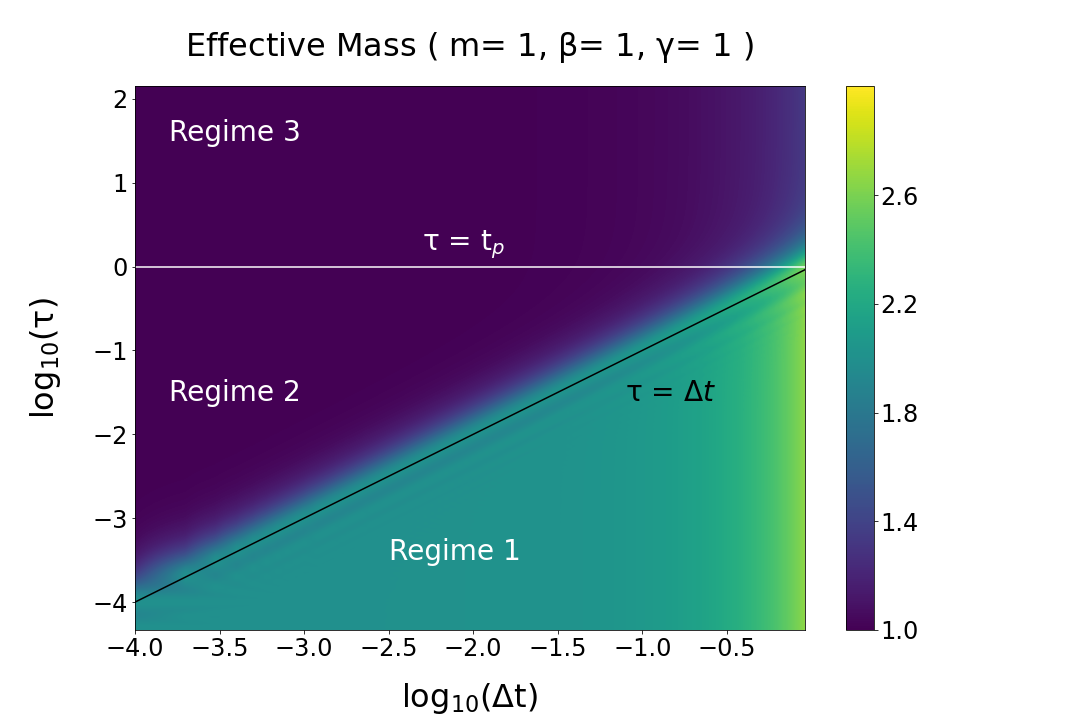}
    \caption{Color contour plot of $m^*$ against the measurement  interval $\Delta t$ and oscillation period $\tau$, with $m = \beta = \gamma =1$.}
\label{fig:Contour}
\end{figure}

Fig.~\ref{fig:Contour} plots $m^*$, given by Eq.~\ref{eq:mstar_Exact_Same}, as a function of $\Delta t$ and $\tau$, at $t_p=1$ and $m=1$. We see agreement with Eq.~\ref{eq:mstar_Regime_Same}: $m^* \approx 2m $ in regime 1, and $m^* \approx m$ in regimes 2 and 3. 

Eq.~\ref{eq:regime1} illustrates that even if the experimental time resolution is adequate to observe a Brownian particle's ballistic motion, i.e\ $\Delta t \ll t_p$, the measured velocity distribution $P(\bar{v}_1)$ may still fail to recover the instantaneous velocity distribution $P^\textrm{MB}(v_1)$, if there exists an additional relevant timescale, such as $\tau$ in our model, that is shorter than $\Delta t$.
Regime 1 is reflected in the experimental situation of Ref.~\cite{Mo2015}, where the time resolution $\Delta t$ is shorter than the momentum relaxation timescale $t_p$, but longer than the response time $\tau_{\textrm{fluid}}$ of the surrounding liquid, resulting in the effective mass given by Eq.~\ref{eq:mstar}.

\section{Brownian Particles with Different Masses}
\label{sec:differentMasses}

We now imagine that the coupled particles have different masses, and we replace Eq.~\ref{eq:Eq_of_Motion_Same} by
\begin{subequations}
\label{eq:Eq_of_Motion_Diff}
    \begin{align}
    m_1\ddot{q}_1 &= -k(q_1-q_2) - \gamma \dot{q}_1 + \sqrt{2\gamma/\beta}\,\xi_1  \\
    m_2\ddot{q}_2 &= -k(q_2-q_1) - \gamma \dot{q}_2 + \sqrt{2\gamma/\beta}\,\xi_2 \quad.
    \end{align}
\end{subequations}
Unlike in the previous section (see Eq.~\ref{eq:Decoupled_Eq_of_Motion}), the equations of motion do not decouple upon transforming to the center of mass and separation variables. Nonetheless, assuming the initial velocities $v_{10}$ and $v_{20}$ and the initial separation $q_0 = q_{10}-q_{20}$ are sampled from equilibrium, we can still solve for the distribution $P_t(q_1|q_{10})$ and ultimately for $P_{\Delta t}(\bar{v}_1)$. Leaving the detailed calculation to the Appendix, we again find an empirical velocity distribution of the form

\begin{align}
    P_{\Delta t}(\bar{v}_1)  = \sqrt{\frac{\beta m^*}{2\pi}}\exp\Big(- \frac{\beta m^*}{2} \bar{v}_1^2\Big)     \label{eq:P(Vbar;Deltat)_Diff}
\end{align}
with
\begin{align}
\label{eq:mstar_Exact_Diff}
    m^* = \frac{\Delta t^2}{\beta}\Big(\sigma_{q_1}^2 + \frac{b^2}{\beta k} + \frac{c^2}{\beta m_1} + \frac{d^2}{\beta m_2} \Big)^{-1} \quad ,
\end{align}
where the expression for $\sigma^2_{q_1}$, $b$, $c$ and $d$ are given by Eqs.~\ref{eq:First_Row_of_Exp_Matrix} and \ref{eq:2nd_Moment_for_q_1_Diff} in the Appendix.
Eq.~\ref{eq:mstar_Exact_Diff} gives a complicated but exact expression for the effective mass $m^*$, which simplifies when there is a separation of timescales.
We introduce
\begin{equation}
    t_{p_i} = \frac{m_i}{\gamma} \quad,\quad \tau = 2\pi\sqrt{\frac{m_1m_2}{k(m_1+m_2)}}
\end{equation}
with $i \in \{1,2\}$.
$t_{p_1}$ and $t_{p_2}$ are the momentum relaxation times for the two particles, which we assume to be comparable: $t_{p_1}\cong t_{p_2}$.
As before, $\tau$ denotes the harmonic oscillation period for the particle separation $q$.
We then find that
\begin{equation}
\label{eq:mstar_Approx_Diff}
    m^* \approx \frac{(m_1+m_2)2\pi^2(\Delta t/\tau)^2}{2\pi^2 (\Delta t/\tau)^2+(m_2/m_1)(1-\cos(2 \pi \Delta t/\tau))}
\end{equation}
when
\begin{equation}
    \tau,\Delta t \ll t_{p_1}, t_{p_2} \quad .
\end{equation}
From Eq.~\ref{eq:mstar_Approx_Diff}, it is straightforward to verify that if we additionally have a separation of timescales between $\Delta t$ and $\tau$, then $m^*$ reduces to approximately $m_1+m_2$ or $m_1$, analogously to regimes 1 and 2 in Eq.~\ref{eq:mstar_Regime_Same}:
\begin{subequations}
\label{eq:mstar_Regime_Diff}
\begin{align}
\label{eq:mstar_Regime1&2_Diff}
    \textrm{regime 1} &: \tau \ll \Delta t \ll t_{p_1} , t_{p_2} \to m^* \approx m_1 + m_2 \\
    \textrm{regime 2} &: \Delta t \ll \tau \ll t_{p_1} , t_{p_2} \to m^* \approx m_1 \quad .
\end{align}
For regime 3, we are unable to obtain a simple approximate expression for $m^*$ analytically, but the numerical evaluation of the exact expression of $m^*$, Eq.~\ref{eq:mstar_Exact_Diff}, suggests
\begin{align}
\label{eq:mstar_Regime3_Diff}
    \textrm{regime 3} &: \Delta t \ll t_{p_1} , t_{p_2} \ll \tau \to m^* \approx m_1  \quad .
\end{align}
\end{subequations}
As in the case of identical masses, if $\Delta t$ is the shortest timescale (regimes 2 and 3), then the instantaneous velocity can be resolved experimentally, and the effective mass is the particle's actual mass; whereas if $\tau$ is the shortest timescale (regime 1), corresponding to a large spring stiffness $k$, the two particles seem to move as a single particle of mass $m_1+m_2$.

\begin{figure}[!hbtp] 
    \includegraphics[width=\columnwidth]{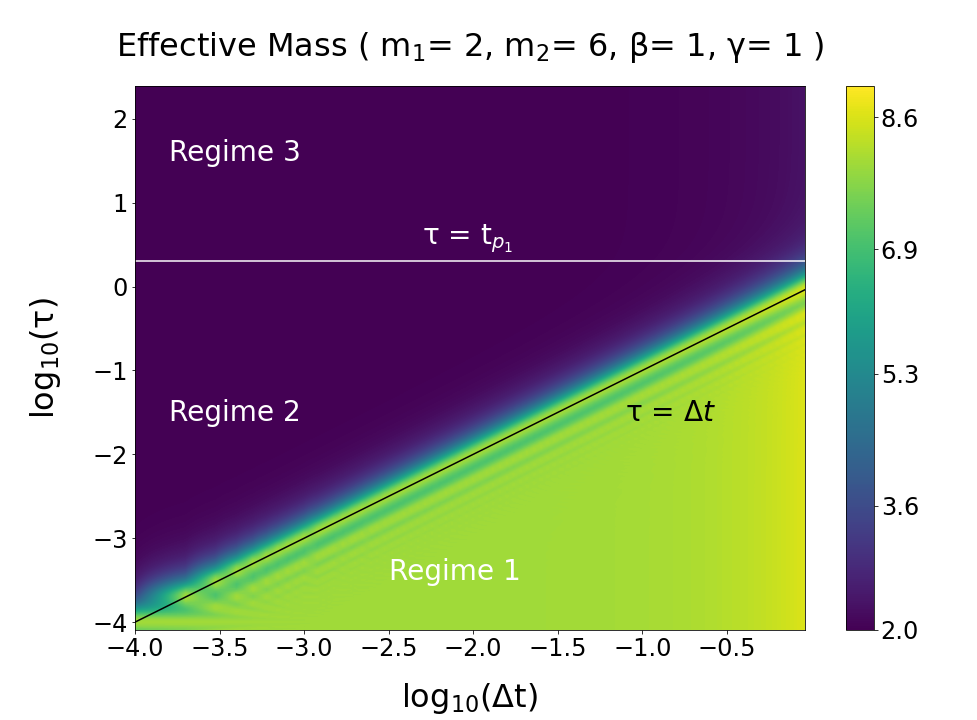}
    \caption{Color contour plot of $m^*$ against the measurement  interval $\Delta t$ and oscillation period $\tau$, for $m_1=2$, $m_2=6$, $\beta=\gamma=1$. }
    \label{fig:Contour_Diff}
\end{figure}

Fig.~\ref{fig:Contour_Diff} plots $m^*$, given by Eq.~\ref{eq:mstar_Exact_Diff}, as a function of $\Delta t$ and $\tau$ with $\gamma =1$, $m_1=2$, and $m_2=6$.
We see agreement with Eq.~\ref{eq:mstar_Regime_Diff}: $m^* \approx m_1 + m_2 $ in regime 1, and $m^* \approx m_1$ in regimes 2 and 3. This behavior is qualitatively similar to that of the case of identical masses.

\section{Numerical Simulations}
We have performed numerical simulations of our model using the Euler-Maruyama method \cite{Maruyama1955}, for different values of $m_1$, $m_2$, $k$ and $\Delta t$, with fixed $\beta=\gamma=1$. To obtain the measured velocity distribution $P_{\Delta t}(\bar{v}_1)$, for every choice of parameters $(m_1,m_2,k)$ we generated $10^5$ trajectories of total duration $t_{\textrm{traj}} = 1$ with a numerical integration time step $\delta t = 10^{-9}$. We then computed
\begin{align}
    \label{eq:numerical-v-bar}
    \bar{v}_1(\Delta t) = \frac{q_1(\Delta t) - q_1(0)}{\Delta t}
\end{align}
for each trajectory and from these values we constructed the distribution $P_{\Delta t}(\bar{v}_1)$.

\begin{figure}[!hbtp] 
    \includegraphics[width=\columnwidth]{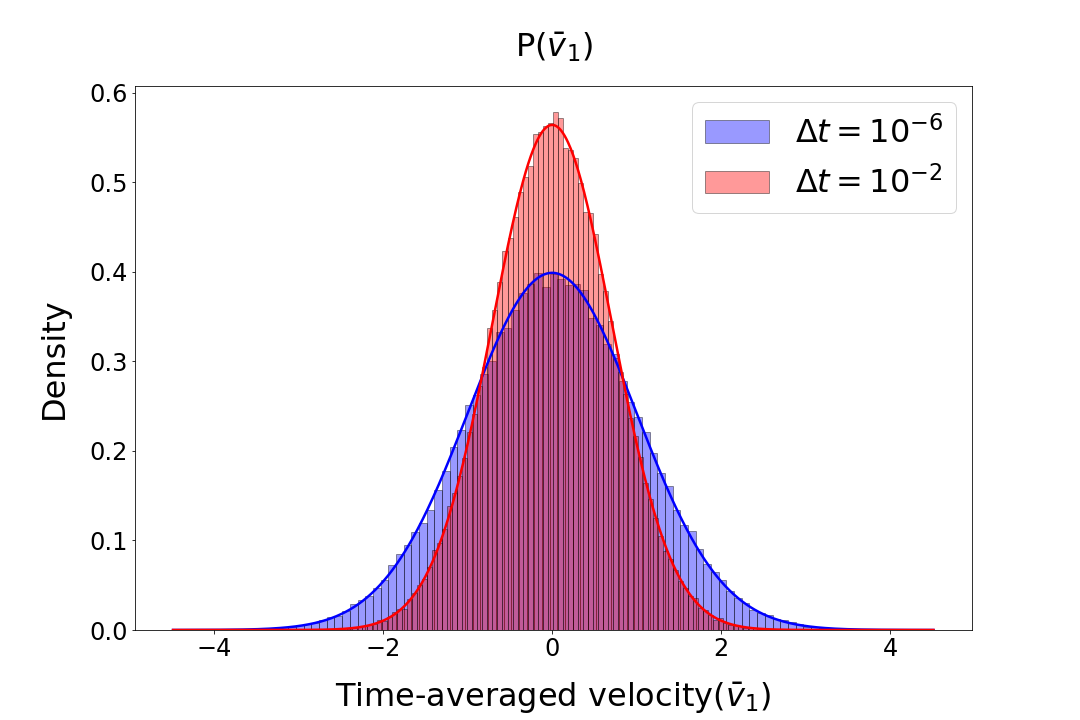}
    \caption{Measured velocity distribution $P(\bar{v}_1)$ with $t_p = 1$, $\tau = 10^{-4}$ and  $m_1 = m_2 =\beta = \gamma = 1$. Red: $\Delta t =10^{-2}$ , Blue: $\Delta t =10^{-6}$.}
    \label{fig:Vbar_Dist}
\end{figure}

Fig.~\ref{fig:Vbar_Dist} shows the measured velocity distribution $P_{\Delta t}(\bar{v}_1)$ obtained from simulations in which both particles have mass $m=1$, with other parameters chosen so that $t_p = 1$ and $\tau = 10^{-4}$. The red and blue histograms correspond to  $P_{\Delta t}(\bar{v}_1)$ with  $\Delta t = 10^{-2}$ (regime 1) and $10^{-6}$ (regime 2) respectively. The solid red and blue curves are zero-mean Gaussians with variances 1/2 and 1, corresponding to effective masses $m^*=2$ and $m^* = 1$, respectively. The numerically obtained distributions agree with the theoretical predictions of Eq.~\ref{eq:mstar_Regime_Same}.

\begin{figure}[!hbtp] \label{fig:m*_vs_Deltat}
\subfigure[Effective mass $m^*$ against the measurement interval $\Delta t$ with $t_p = 1$ and $\tau = 10^{-4}$  $(m =  \beta = \gamma = 1)$.]{\includegraphics[width=\columnwidth]{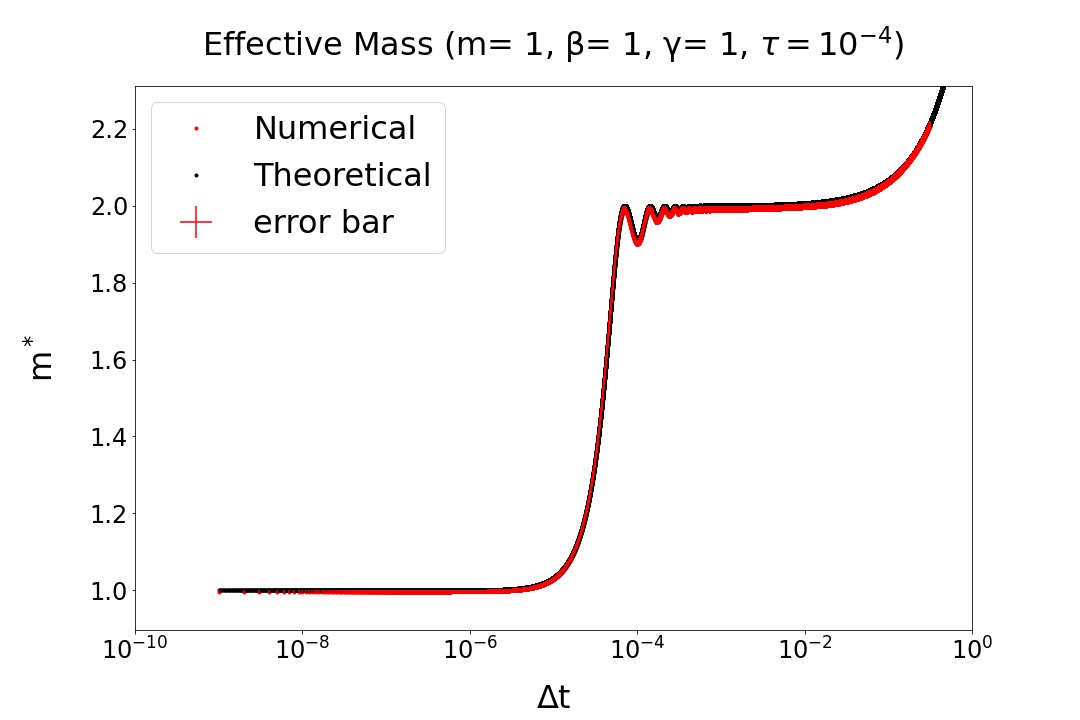}\label{fig:m*vsDeltat_Same}}\\

\subfigure[Effective mass $m^*$ against the measurement interval $\Delta t$ with $\tau = 10^{-4}$  $(m_1 = 2 , m_2 =6,  \beta = \gamma  = 1)$.]{\includegraphics[width=\columnwidth]{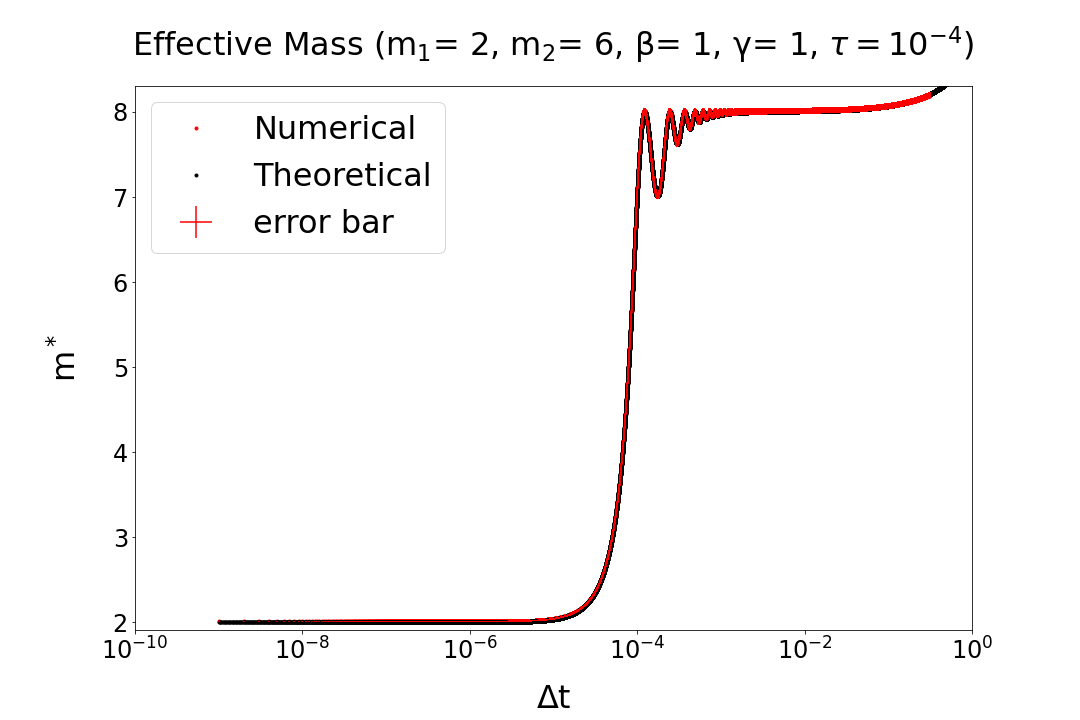}\label{fig:m*vsDeltat_Diff}}\\

\caption{Effective mass $m^*$ against the measurement interval $\Delta t$}
\label{fig:m*vsDeltat}
\end{figure}

Fig.~\ref{fig:m*vsDeltat} shows how $m^*$ varies with $\Delta t$, at a fixed $\tau$ and $t_p = m/\gamma$ (or $t_{p_i} = m_i/\gamma$ when the masses differ). 
The red points are values of $m^*$ obtained from simulations, while the black curves show the analytical predictions of Eq.~\ref{eq:mstar_Exact_Same} (Fig.~\ref{fig:m*vsDeltat_Same}) and Eq.~\ref{eq:mstar_Exact_Diff} (Fig.~\ref{fig:m*vsDeltat_Diff}).
We observe excellent agreement between simulation results and analytical predictions.
Both figures show $m^*\approx m_1$ at small values of $\Delta t$, along with a transition around $\Delta t = \tau$ to a plateau $m^*\approx m_1+m_2$, corresponding to a transition from regime 2 to regime 1 as predicted by Eqs.~\ref{eq:mstar_Regime_Same} and \ref{eq:mstar_Regime_Diff}. 

Notice the wiggles in Fig.~\ref{fig:m*vsDeltat} at $\Delta t \approx \tau$. Mathematically, from Eq.~\ref{eq:mstar_Approx_Diff}, which is valid as long as $\Delta t,\tau \ll t_p \,\, (t_{p_i})$, these wiggles arise from the cosine appearing in the denominator. In regime 1, where $\Delta t/\tau \gg 1$, the term $2\pi^2(\Delta t/\tau)^2$ in the denominator of Eq.~\ref{eq:mstar_Approx_Diff} dominates over the cosine term, masking the latter's oscillations. In regimes 2 and 3, where $\Delta t /\tau \ll 1$, if we express $m^*$ as a power series in $\Delta t / \tau$, we obtain
\begin{align}
    m^* = m_1 + \frac{m_1 \pi^2}{3(m_1+m_1)}\Big(\frac{\Delta t}{\tau}\Big)^2 + O\Big(\Big(\frac{\Delta t}{\tau}\Big)^4\Big)
\end{align}
which increases monotonically with $\Delta t$. Therefore, we do not see wiggles whenever a time separation between $\Delta t$ and $\tau$ exists. However, when $\Delta t$ and $\tau$ are comparable, the cosine term's oscillatory nature becomes significant. In fact, the crests and troughs correspond to integer and half-integer values of $\Delta t/\tau$, suggesting that the wiggles in $m^*$ at $\Delta t \approx \tau$ arises from synchronization between the measurements and the oscillation of the particles.

Also note that in Fig.~\ref{fig:m*vsDeltat} at $\Delta t \approx t_p \,\, (t_{p_i})$, the value of $m^*$ increases with $\Delta t$. This growing tail is expected because for $t_p \,\,(t_{p_i}) \centernot{\gg} \Delta t$, the observed dynamics are no longer ballistic but diffusive. In a diffusion process, the variance of the displacement $\Delta q_1$ scales linearly with the time interval $\Delta t$. As a result, the variance of the time-averaged velocity $\bar{v}_1$ scales as $\Delta t^{-1}$, and thus $m^*$ scales as $\Delta t$, leading to the exponential growth observed in the logarithmic scale in Fig.~\ref{fig:m*vsDeltat}.

\section{Summary}
As discussed in the Introduction, the effective mass of a Brownian sphere in a fluid ranges from $m^*\approx m$ to $m^*\approx m + (1/2)M_d$, depending on how the measurement time resolution compares with the fluid's hydrodynamic response time. Modeling this behavior with a pair of harmonically coupled, underdamped Brownian particles, we have solved exactly for the effective mass, $m^*$, in terms of the actual mass, $m$, and three relevant timescales: the momentum relaxation time, $t_p$, the harmonic oscillation period, $\tau$, and the measurement time interval, $\Delta t$ (Eq.~\ref{eq:mstar_Exact_Same}).  When these timescales are widely separated, the effective mass simplifies (Eq.~\ref{eq:mstar_Regime_Same}). We find $m^*\approx m$ when $\Delta t$ is the shortest timescale, in other words when position measurements are sufficiently frequent to resolve the particle's instantaneous velocity. However, if $\tau\ll\Delta t$, then these measurements do not capture the rapid oscillations due to stiff harmonic coupling; the particles then appear to move as if glued together: $m^*\approx 2m$.
These results generalize to the case when the particles have different masses (Eqs.~\ref{eq:mstar_Exact_Diff}, \ref{eq:mstar_Regime_Diff}).
We have also presented the results of numerical simulations, verifying our analytical calculations.

\section{Acknowledgments}
This research was supported by the U.S. National Science Foundation under Grant No. 2127900.
CJ acknowledges stimulating discussions with Mark Raizen and Kanupriya Sinha.

\bibliography{Reference.bib}

\onecolumngrid
\section {Appendix}

\setcounter{equation}{0}
\renewcommand{\theequation}{A\arabic{equation}}
\subsection{Appendix A : $P_t(q_1|q_{10})$ for Coupled Brownian Particles with Identical Masses}
We first rewrite Eqs.~\ref{eq:Decoupled_Eq_of_Motion_for_Q} and \ref{eq:Decoupled_Eq_of_Motion_for_V} as follows:
\begin{equation} 
    \frac{d}{dt} X(t)  =  -\Lambda X(t) + F    , \label{eq:Decoupled_Eq_of_Motion_Linear}
\end{equation}
with
\begin{align}
    X(t) = \begin{pmatrix} Q\\V \end{pmatrix} \quad, \quad \Lambda = \begin{pmatrix} 0&1 \\ 0& \gamma/m\end{pmatrix}  \quad , \quad F(t) =  \frac{\sqrt{2\gamma/\beta}}{2m}\begin{pmatrix} 0 \\ \xi_1 +\xi_2 \end{pmatrix} \quad .
\end{align}
The general solution of Eq.~\ref{eq:Decoupled_Eq_of_Motion_Linear} is:
\begin{align} 
\label{eq:Gen_Sol_for_Eq_of_Motion_Same}
    X(t) &= e^{-\Lambda t}X(0) + \int^{t}_{0} dt\,'\, e^{-\Lambda (t-t\,')}F(t\,') \\
    \nonumber \\ 
    e^{-\Lambda t} &= \begin{pmatrix} 1 & \frac{m}{\gamma}(1-e^{-\gamma t/m}) \\ 0 & e^{-\gamma t/m} \end{pmatrix} \quad .
\end{align}
From this solution we obtain
\begin{align} 
\label{eq:Exact_Sol_for_Q&V}
    Q(t) & =  Q(0) + \frac{m}{\gamma}(1-e^{-\gamma t/m})V(0) + \frac{\sqrt{2\gamma/\beta}}{2m}\int^{t}_{0}\, dt' \frac{m}{\gamma}\,\Big(1-e^{-\gamma (t-t')/m}\Big)\,(\xi_1(t') +\xi_2(t'))\\
     \nonumber \\
    V(t) & =  e^{-\gamma t/m}V(0) + \frac{\sqrt{2\gamma/\beta}}{2m}\int^{t}_{0}\, dt' e^{-\gamma (t-t')/m}(\xi_1(t') +\xi_2(t')) \quad .
\end{align}
Taking the ensemble average for both $Q(t)$ and $V(t)$, the integral terms vanish since $\xi_1$ and $\xi_2$ are zero-mean Gaussian white noise, and we have
\begin{align} 
\label{eq:Mean_for_Q}
    \langle Q \rangle  & =  Q(0) + \frac{m}{\gamma}(1-e^{-\gamma t/m})V(0) \\
    \nonumber \\
    \langle V \rangle & =  e^{-\gamma t/m}V(0) \label{eq:Mean_for_V} \quad .
\end{align}
Combining Eq.~\ref{eq:Gaussian_White_Noise} and Eqs.~\ref{eq:Exact_Sol_for_Q&V} - \ref{eq:Mean_for_V}, we then compute the variances and the covariance:
\begin{align} 
\label{eq:2nd_Moment_for_Q&V}
    \sigma^2_\textrm{QQ} & = \langle (Q-\langle Q \rangle)^2\rangle = \frac{1}{2\gamma\beta}\int^t_0 dt'\Big(1-e^{-\gamma (t-t')/m}\Big)^2 = \frac{m}{2\beta\gamma^2} \Big( 2 \frac{\gamma}{m}t - e^{-2\gamma t/m} +4e^{-\gamma t/m} - 3 \Big) \\
    \nonumber\\
    \sigma^{2}_\textrm{QV} & = \langle (Q-\langle Q \rangle)(V-\langle V \rangle)\rangle = \frac{1}{2\beta m} \int^t_0 dt' \Big(1-e^{-\gamma (t-t')/m}\Big) e^{-\gamma(t-t')/m} =  \frac{1}{2\beta\gamma} \big( 1-e^{-\gamma t/m} \big)^{2} \\
    \nonumber \\
    \sigma^{2}_\textrm{VV} & = \langle (V-\langle V \rangle)^2\rangle = \frac{\gamma}{2\beta m^2} \int^t_0 dt' e^{-2\gamma(t-t')/m} = \frac{1}{2\beta m} \big( 1-e^{-2\gamma t/m} \big) \quad .
\end{align}

Applying the same procedure to Eqs.~\ref{eq:Decoupled_Eq_of_Motion_for_q} and \ref{eq:Decoupled_Eq_of_Motion_for_v}, we obtain
\begin{align} 
    \langle q \rangle & =  \frac{\big(\lambda_+e^{-\lambda_-t}-\lambda_-e^{-\lambda_+t}\big)q(0)+\big(e^{-\lambda_-t}-e^{-\lambda_+t}\big)v(0)}{\lambda_+-\lambda_-} \\
     \nonumber\\
    \langle v \rangle & =  \frac{\omega^2\big(e^{-\lambda_+t}-e^{-\lambda_-t}\big)q(0) + \big(\lambda_+e^{-\lambda_+t}-\lambda_-e^{-\lambda_-t}\big)v(0)}{\lambda_+ - \lambda_-} \\
     \nonumber\\
    \sigma^{2}_\textrm{qq} & =  \frac{2\gamma\Big(\frac{\gamma}{m\omega^2}-\frac{4m}{\gamma}\big(1-e^{-\gamma t/m}\big) -\frac{e^{-2\lambda_-t}}{\lambda_-}-\frac{e^{-2\lambda_+t}}{\lambda_+}\Big)}{\beta m^2(\lambda_+-\lambda_-)^2}\\
     \nonumber \\
    \sigma^2_\textrm{qv} & =  \frac{2\gamma\Big(e^{-\lambda_+t}-e^{-\lambda_-t}\Big)^2}{\beta m^2(\lambda_+-\lambda_-)^2}  \\
     \nonumber \\
    \sigma^2_\textrm{vv} & =  \frac{2\gamma\Big(\frac{\gamma}{m}-\frac{4m\omega^2}{\gamma}(1-e^{-\gamma t/m})-\lambda_-e^{-2\lambda_-t}-\lambda_+e^{-2\lambda_+t}\Big) }{\beta m^2(\lambda_+-\lambda_-)^2}\label{eq:1st&2nd_Moment_for_q&v} ,
\end{align}
with
\begin{equation}
    \omega^2 = \frac{2k}{m} \quad,\quad \lambda_\pm  = \frac{1}{2}\Bigl(\frac{\gamma}{m} \pm \sqrt{\frac{\gamma^2}{m^2}-4\omega^2}\Bigr) .   
\end{equation}

The vectors $\mathbf{X}$ and $\mathbf{x}$ and matrices $\mathbf{C}$ and $\mathbf{c}$ appearing in the conditional probability distributions $P_t(Q,V|Q_0,V_0)$ and $P_t(q,v|q_0,v_0)$, Eq.~\ref{eq:P(Q,V|Q_0,V_0)&P(q,v|q_0,v_0)=Gaussian}, are
\begin{align} 
    \mathbf{X} &= \begin{pmatrix} Q - \langle Q \rangle \\ V - \langle V \rangle \end{pmatrix} \quad , \quad \mathbf{x} = \begin{pmatrix} q - \langle q \rangle \\ v - \langle v \rangle \end{pmatrix} \\
    \nonumber\\
    \mathbf{C} &= \begin{pmatrix} \sigma^2_\textrm{QQ} & \sigma^2_\textrm{QV} \\ \sigma^2_\textrm{QV} & \sigma^2_\textrm{VV} \end{pmatrix}  \quad , \quad \mathbf{c} = \begin{pmatrix} \sigma^2_\textrm{qq} & \sigma^2_\textrm{qv} \\ \sigma^2_\textrm{qv} & \sigma^2_\textrm{vv} \end{pmatrix} \label{eq:covariance-matrix-of-qv-appendix}
\end{align}
with the first and second moments of $(Q,V)$ and $(q,v)$ given by Eqs.~\ref{eq:Mean_for_Q} - \ref{eq:1st&2nd_Moment_for_q&v}.
Marginalizing the conditional distributions yields
\begin{align} 
    \label{eq:P(Q|Q_0,V_0)}
    P_t(Q|Q_0,V_0) & = \int^{\infty}_{-\infty}\,dV P_t(Q,V|Q_0,V_0) = \sqrt{\frac{1}{2\pi \sigma^2_\textrm{QQ}}}\exp\Big(\frac{1}{2\sigma^2_\textrm{QQ}}\big(Q-\langle Q \rangle\big)^2\Big) \\
    \nonumber\\
    \label{eq:P(V|Q_0,V_0)&P(v|q0,v0)}
    P_t(V|Q_0,V_0) & = \int^{\infty}_{-\infty}\,dQ P_t(Q,V|Q_0,V_0) = \sqrt{\frac{1}{2\pi \sigma^2_\textrm{VV}}}\exp\Big(\frac{1}{2\sigma^2_\textrm{VV}}\big(V-\langle V \rangle\big)^2\Big) \\
     \nonumber \\
    P_t(q|q_0,v_0) & =  \int^{\infty}_{-\infty}\,dvP_t(q,v|q_0,v_0)  =  \sqrt{\frac{1}{2\pi \sigma^2_\textrm{qq}}}\exp\Big(\frac{1}{2\sigma^2_\textrm{qq}}\big(q-\langle q \rangle\big)^2\Big) \label{eq:P(q|q0,v0)} \\
    \nonumber\\
    P_t(v|q_0,v_0) & =  \int^{\infty}_{-\infty}\,dqP_t(q,v|q_0,v_0) =  \sqrt{\frac{1}{2\pi \sigma^2_\textrm{vv}}}\exp\Big(\frac{1}{2\sigma^2_\textrm{vv}}\big(v-\langle v \rangle\big)^2\Big) \quad .
\end{align}
From these expressions, we see that in long-time limit $t \to \infty$, the variables $V$, $v$, and $q$ settle to the equilibrium distributions:
\begin{align} 
     P_\textrm{eq}(V) & =  \lim_{t \to \infty} P(V,t|Q_0,V_0)  = \sqrt{\frac{\beta m}{\pi}}e^{-\beta m V^2}  \label{eq:P_eq(V)}    \\
    \nonumber\\
    P_\textrm{eq}(v) & =  \lim_{t \to \infty} P(v,t|q_0,v_0)  = \sqrt{\frac{\beta m}{4\pi}}e^{-\beta m v^2/4} \label{eq:P_eq(v)}\\
    \nonumber \\
     P_\textrm{eq}(q) &= \lim_{t \to \infty}P_t(q|q_0,v_0) = \sqrt{\frac{\beta m \omega^2}{4\pi}} e^{-\beta m \omega^2 q^2/4} \label{eq:P_eq(q)}\quad .
\end{align}

Assuming the initial velocities are drawn from the equilibrium distributions Eqs.~\ref{eq:P_eq(V)} and \ref{eq:P_eq(v)}, we obtain $P_t(Q|Q_0)$ and $P_t(q|q_0)$ by integrating out the dependence on $V_0$ and $v_0$ from Eqs.~\ref{eq:P(Q|Q_0,V_0)} and \ref{eq:P(q|q0,v0)} respectively.

\begin{align}
    \label{eq:P(Q|Q_0)}
        P_t(Q|Q_0) & =  \int^{\infty}_{-\infty}\,dV_0P_t(Q|Q_0,V_0)P_\textrm{eq}(V_0) =\frac{1}{\sqrt{2\pi\sigma^2_{Q}} }\exp\Big(-\frac{1}{2\sigma^2_Q}\big(Q - \bar{Q}\big)^2\Big) \\
        \nonumber\\
        P_t(q|q_0) & =  \int^{\infty}_{-\infty}\,dv_0P_t(q|q_0,v_0)P_\textrm{eq}(v_0) =\frac{1}{\sqrt{2\pi\sigma^2_{q}} }\exp\Big(-\frac{1}{2\sigma^2_q}\big(q - \bar{q}\big)^2\Big) \label{eq:P(q|q_0)}
\end{align}
\begin{align}
        \bar{Q} &= Q_0 \quad,\quad \sigma^2_{Q} = \frac{1}{\beta \gamma} \Big( t - \frac{m}{\gamma}+\frac{m}{\gamma}e^{-\gamma t /m} \Big) \\
        \nonumber \\
        \bar{q} &= \alpha q_0 \quad,\quad \sigma^2_{q} = \frac{2}{\beta m \omega^2}\big(1-\alpha^2\big) \quad,\quad  \alpha  = \frac{\lambda_+ e^{-\lambda_-t}-\lambda_-e^{-\lambda_+t}}{\lambda_+ - \lambda_-} \quad .
\end{align}

Note that both $P_t(Q|Q_0)$ and $P_t(q|q_0)$ are Gaussians. Since $Q$ and $q$ are independence random variables and $q_1 = Q + q/2$, it follows that $P_t(q_1|Q_0,q_0)$ is also a Gaussian:
\begin{equation}
\label{eq:P(q1|Q_0,q_0)_appendix}
    P_t(q_1|Q_0,q_0) = \frac{1}{\sqrt{2\pi\sigma^2_{q_1}}} \exp\Big( -\frac{1}{2\sigma^2_{q_1}}\big(q_1 - \bar{q}_1 \big)^2 \Big) 
\end{equation}
with mean $\bar{q}_1 = \bar{Q}+\bar{q}/2$ and variance $\sigma^2_{q_1} = \sigma^2_{Q}+ \sigma^2_{q}/4$.

Finally, assuming the initial separation $q_0$ to be sampled from the equilibrium distribution Eq.~\ref{eq:P_eq(q)}, we obtain $P_t(q_1|q_{10})$ from Eq.~\ref{eq:P(q1|Q_0,q_0)_appendix} by first performing a change of variables, using $Q_0 = q_{10} -q_0/2$, and then integrating out the dependence on $q_0$:
\begin{equation} 
\label{eq:P(q1|q_10,q_0)_Appendix}
       P_t(q_1|q_{10},q_0) = \int dQ_0 \, P_t(q_1|Q_0,q_0)\,\delta\big(Q_0 - q_{10}+\frac{1}{2}q_0\big)
\end{equation}
\begin{align} 
    P_t(q_1|q_{10}) &= \int^{\infty}_{-\infty}\,dq_0 \, P_t(q_1|q_{10},q_0)P_\textrm{eq}(q_0) = \frac{1}{\sqrt{2\pi \sigma^2(t)}}\exp\Big(-\frac{1}{2\sigma^2(t)}\big(q_1-q_{10}\big)^2\Big)
\end{align}
\begin{align} 
    \label{eq:P(q_1|q_10)_Appendix and sigma_Exact_Same_Appendix}
        \sigma^2(t) = \frac{t}{\beta \gamma}+\frac{1}{2\beta m \omega^2} \Big\{ 2 - \frac{2m^2\omega^2}{\gamma^2}\big( 1-e^{-\gamma t/m} \big) - 2e^{-\gamma t/2m}\Big[\frac{\gamma}{ma}\sinh\big(\frac{at}{2}\big)+\cosh\big(\frac{at}{2}\big)\Big]\Big\} \quad , \quad 
        a = \sqrt{\frac{\gamma^2}{m^2}-4\omega^2}
\end{align}

\setcounter{equation}{0}
\renewcommand{\theequation}{B\arabic{equation}}
\subsection{Appendix B : Approximate Expression of $m^*$ for Coupled Brownian Particles with Identical Masses}
In Eq. ~\ref{eq:mstar_Regime_Same}, the timescale separations $t_p\gg \tau$ and $t_p\gg \Delta t$ are valid in regimes 1 and 2, implying $\gamma/m\omega \ll 1$ and $\gamma \Delta t/m \ll 1$. To leading order in $\gamma/m\omega$ and $\gamma \Delta t/m$, Eqs.~\ref{eq:sigma_Exact_Same} and \ref{eq:mstar_Exact_Same} become:
\begin{equation}
    \label{eq:sigma_Approx_Regime1&2_Same}
        \sigma_{\Delta q_1}^2(\Delta t)  \approx  \frac{1}{2\beta m \omega^2} \Big( 2 + \omega^2 \Delta t^2 - 2\cos\big(\omega \Delta t \big) \Big)
\end{equation} 
\begin{equation}
    \label{eq:mstar_Approx_Regime1&2_Same}
    m^*  \approx  2m \Big[ \frac{\omega^2 \Delta t^2}{\omega^2 \Delta t^2 + 2 - 2\cos(\omega \Delta t)} \Big] \quad .
\end{equation}

For regime 1, we also have $\Delta t \gg \tau$, i.e.\ $\omega \Delta t \gg 1$; while in regime 2, we have  $\tau \gg \Delta t$, hence $\omega \Delta t \ll 1$. Therefore, from Eq.~\ref{eq:mstar_Approx_Regime1&2_Same}, we can further approximate $m^*$:
\begin{align} 
\label{eq:mstar_Regime1&2_Same_Appendix}
    \textrm{regime }1 &: m^* \approx 2m\Big(\frac{\omega^2 \Delta t^2}{\omega^2\Delta t^2}\Big)=2m \\
    \nonumber \\
    \textrm{regime } 2 &: m^* \approx 2m\Big(\frac{\omega^2\Delta t^2}{\omega^2 \Delta t^2+2-2(1-\omega^2 \Delta t^2/2)}\Big)=m \quad .
\end{align}

The timescale separations that define regime 3 imply $\gamma \Delta t/m \ll 1$, $m\omega/\gamma \ll 1$, and $\omega \Delta t\ll 1$. Hence, to leading order in $\gamma \Delta t/m$ and $m\omega^2 \Delta t/ \gamma$, Eq.~\ref{eq:sigma_Exact_Same} gives
\begin{equation}
\label{eq:sigma_Approx_Regime3_Same}
        \sigma_{\Delta q_1}^2(\Delta t) \approx \frac{\Delta t^2}{\beta m}\quad .
\end{equation}
Therefore, Eq.~\ref{eq:mstar-in-terms-of-variance} yields
\begin{equation}
\label{eq:mstar_Approx_Regime3_Same}
    \textrm{regime} 3 : \,\,m^* =\frac{1}{\beta \sigma^2_{\bar{v}_1}} = \frac{\Delta t^2}{\beta  \sigma_{\Delta q_1}^2(\Delta t)}\approx \frac{\Delta t^2}{\beta} \Big(\frac{\beta m}{\Delta t^2} \Big) = m \quad .
\end{equation}

\setcounter{equation}{0}
\renewcommand{\theequation}{C\arabic{equation}}
\subsection{Appendix C : $P_t(q_1|q_{10})$ and Approximate Expression of $m^*$ for Coupled Brownian Particles with Different Masses}
We first rewrite Eq.~\ref{eq:Eq_of_Motion_Diff} as follows:
\begin{align} 
    \label{eq:Eq_of_Motion_Diff_Linear}
    \frac{d}{dt} X(t)  =  \Lambda'\,X(t) + F\,' 
\end{align}
\begin{align}
    X(t) = \begin{pmatrix} q_1 \\q_2\\v_1\\v_2\\\end{pmatrix} \quad , \quad \Lambda' =  \begin{pmatrix}
    0 & 0 & 1 & 0\\
    0 & 0 & 0 & 1\\
    -k/m_1 & k/m_1 & -\gamma/m_1& 0 \\
    k/m_2 & -k/m_2 & 0 & -\gamma/m_2
    \end{pmatrix} \quad , \quad F'(t) = 
     \sqrt{2\gamma/\beta} \begin{pmatrix} 0 \\0\\ \xi_1/m_1 \\ \xi_2/m_2 \end{pmatrix}\quad .
\end{align}
The general solution for Eq.~\ref{eq:Eq_of_Motion_Diff_Linear} is:
\begin{equation}
\label{eq:Gen_Sol_for_Eq_of_Motion_Diff}
    X(t) = e^{\Lambda't}X(0) + \int^{t}_{0} dt\,'\, e^{\Lambda'(t-t\,')}F\,'(t\,') \quad .
\end{equation}
The matrix exponential $e^{\Lambda' t}$ in Eq.~\ref{eq:Gen_Sol_for_Eq_of_Motion_Diff} is a $4\times 4$ matrix whose first-row elements are:
\begin{align}
\label{eq:First_Row_of_Exp_Matrix}
    e^{\Lambda'\,t}(1,1)= a(t) = 1/2 + \sum^{3}_{i=1}\frac{\gamma+m_1\lambda_i}{m_1m_2}A_iB_ie^{\lambda_it} & \quad , \quad 
    e^{\Lambda'\,t}(1,2)  = b(t) = 1/2 - \sum^{3}_{i=1}\frac{\gamma+m_1\lambda_i}{m_1m_2}A_iB_ie^{\lambda_it} \nonumber \\
    \nonumber\\
    e^{\Lambda'\,t}(1,3)  = c(t) = \frac{m_1}{2\gamma} + \sum^{3}_{i=1} \frac{A_iB_i}{m_2}e^{\lambda_it}  &\quad , \quad
     e^{\Lambda'\,t}(1,4)  = d(t) = \frac{m_2}{2\gamma} + \sum^{3}_{i=1} \frac{k A_i}{m_1}e^{\lambda_it} 
\end{align}
where
\begin{align}
    A_1 & = \Big[ \lambda_1 (\lambda_2 - \lambda_1)(\lambda_3 - \lambda_1) \Big]^{-1} \quad , \quad A_2  = \Big[ \lambda_2 (\lambda_1 - \lambda_2)(\lambda_3 - \lambda_2) \Big]^{-1} \quad , \quad  A_3 & = \Big[ \lambda_3 (\lambda_1 - \lambda_3)(\lambda_2 - \lambda_3) \Big]^{-1} 
\end{align}
\begin{align}
        B_i &= m_2\lambda_i^2 + \gamma\lambda_i+k 
\end{align}
\noindent and $\lambda_i$ [$i \in \{1,2,3\}$] are the roots of the cubic equation 
\begin{align}
    x^3 + \left(\frac{\gamma}{m_1} + \frac{\gamma}{m_2}\right)x^2 + \left(\frac{k}{m_1} + \frac{k}{m_2} + \frac{\gamma^2}{m_1m_2}\right)x + \frac{2k\gamma}{m_1m_2} =0 \quad . \label{eq:cubic-eq-for-lambda}
\end{align}

Substituting Eq.~\ref{eq:First_Row_of_Exp_Matrix} into  Eq.~\ref{eq:Gen_Sol_for_Eq_of_Motion_Diff}, we obtain $q_1(t)$ in terms of the initial values of the variables:
\begin{align}
    q_1(t) & = a(t) q_{10} + b(t) q_{20} +c(t) v_{10} + d(t) v_{20} + \int^{t}_0 dt' c(t-t') \frac{\sqrt{2\gamma/\beta}}{m_1}\xi_1(t') + \int^{t}_0 dt' d(t-t') \frac{\sqrt{2\gamma/\beta}}{m_2}\xi_2(t')  \quad .\label{eq:explicit-expression-for-q1}
\end{align}
Taking the ensemble average gives:
\begin{align}
    \langle q_1 \rangle & = a(t) q_{10} + b(t) q_{20} +c(t) v_{10} + d(t) v_{20}  \quad .\label{eq:1st_Moment_for_q_1_Diff}
\end{align}
From Eqs.~\ref{eq:Gaussian_White_Noise}, \ref{eq:explicit-expression-for-q1} and \ref{eq:1st_Moment_for_q_1_Diff}, we obtain the variance of $q_1$:
\begin{align}
    \sigma_{q_1}^2 & = \langle (q_1 - \langle q_1 \rangle)^2 \rangle = \frac{2\gamma}{\beta m_1^2}\int^{t}_{0} c^2(t-t')dt' + \frac{2\gamma}{\beta m_2^2}\int^{t}_{0} d^2(t-t')dt'  \quad .\label{eq:2nd_Moment_for_q_1_Diff}
\end{align}

By Eq.~\ref{eq:Eq_of_Motion_Diff_Linear}, the variables ($q_1$, $q_2$, $v_1$, $v_2$) evolve under linear underdamped Langevin equations with independent Gaussian white noises. Therefore, the conditional probability $P(q_1,q_2,v_1,v_2,t|q_{10},q_{20},v_{10},v_{20})$ is a multivariate Gaussian. Integrating out the dependence on $q_2, v_1,$ and $v_2$, we have the marginal distribution:
\begin{equation} 
\label{eq:P(q_1|q_10,q_20,v_10,v_20)}
    P_t(q_1| q_{10}, q_{20}, v_{10}, v_{20})  = \sqrt{\frac{1}{2\pi\sigma^2_{q_1}}}\exp\Big(\frac{-1}{2\pi\sigma^2_{q_1}} (q_1 - \langle q_1 \rangle)^2 \Big) \quad .
\end{equation}.

Letting $q = q_1 - q_2$ denote the separation between the particles, we have the initial separation $q_{0} = q_{10} - q_{20}$. We then perform a change of variables in Eq.~\ref{eq:P(q_1|q_10,q_20,v_10,v_20)} to obtain
\begin{equation}
\label{eq:P(q_1|q_10,q_0,v_10,v_20)}
     P_t(q_1|q_{10},q_{0},v_{10}, v_{20})  = \int dq_{20} P(q_1,t|q_{10},q_{20},v_{10}, v_{20})\delta(q_{20} - q_{10} + q_0)  \quad .
\end{equation}

Assuming the initial separation and the initial velocities $q_0 , v_{10}$, and $v_{20}$ obey the equilibrium distributions,
\begin{align}
\label{eq:P_eq(q)&P_eq(v_1)&P_eq(v2)_Diff}
    P_{\textrm{eq}}(q_0) & = \sqrt{\frac{\beta k}{2\pi}}\exp \Big( \frac{-\beta}{2} k q_0^2 \Big)\\
    \nonumber\\
    P_{\textrm{eq}}(v_{10}) & = \sqrt{\frac{\beta m_1}{2\pi}}\exp \Big( \frac{-\beta}{2} m_1 v_{10}^2 \Big) \\
    \nonumber\\
    P_{\textrm{eq}}(v_{20}) & = \sqrt{\frac{\beta m_2}{2\pi}}\exp \Big( \frac{-\beta}{2} m_2 v_{20}^2 \Big)  \quad ,
\end{align} 
we integrate out the dependence on $q_0 , v_{10}$, and $v_{20}$ from $P(q_1|q_{10},q_0,v_{10},v_{20})$ to obtain $P_t(q_1|q_{10})$:
\begin{align} 
\label{eq:P(q_1|q_10)_Diff}
    P_t(q_1|q_{10})  & = \int^{\infty}_{-\infty}dq_0 \int^{\infty}_{-\infty}dv_{10} \int^{\infty}_{-\infty}dv_{20} \,\, P(q_1|q_{10},q_0,v_{10},v_{20}) P_{eq}(q_0) P_{eq}(v_{10}) P_{eq}(v_{20})\\
    \nonumber\\
    &= \sqrt{\frac{1}{2\pi\sigma_{\Delta q_1}^2}}\exp\Big( \frac{-1}{2 \sigma_{\Delta q_1}^2} ( q_1 - q_{10} )^2 \Big) \\
\end{align}
with
\begin{align}
    \sigma_{\Delta q_1}^2(t) &= \sigma_{q_1}^2(t) + \frac{b^2(t)}{\beta k} + \frac{c^2(t)}{\beta m_1} + \frac{d^2(t)}{\beta m_2} \label{eq:sigma_Diff}  \quad .
\end{align}

Therefore, by Eq.~\ref{eq:mstar-in-terms-of-variance}, the effective mass $m^*$ is
\begin{align}
\label{eq:mstar_Exact_Diff_appendix}
    m^* = \frac{1}{\beta \sigma^2_{\bar{v}_1}} = \frac{\Delta t^2}{\beta \sigma_{\Delta q_1}^2(\Delta t)} = \frac{\Delta t^2}{\beta}\Big(\sigma_{q_1}^2(\Delta t) + \frac{b^2(\Delta t)}{\beta k} + \frac{c^2(\Delta t)}{\beta m_1} + \frac{d^2(\Delta t)}{\beta m_2} \Big)^{-1} 
\end{align} 

Eq.~\ref{eq:mstar_Exact_Diff_appendix} is an exact expression. In the following, assuming the timescale separations described by Eq.~\ref{eq:mstar_Regime_Diff}, we compute approximate expression for $m^*$. To start, we obtain approximate expressions for the $\lambda_i$'s. Since these are the roots of a cubic equation, Eq.~\ref{eq:cubic-eq-for-lambda}, we have
\begin{align} 
    \lambda_1 & = -\frac{m_1+m_2}{3m_1m_2}\gamma - \frac{2^{1/3}}{3}\frac{\Delta_1}{\big(\Delta_2+\sqrt{\Delta_2^2+4\Delta_1^3}\Big)^{1/3}}+\frac{1}{3\cdot2^{1/3}}\big( \Delta_2 + \sqrt{\Delta_2^2+4\Delta_1^3}\big)^{1/3}  \label{eq:lambda-1-exact}\\
    \nonumber \\
    \lambda_2 & = -\frac{m_1+m_2}{3m_1m_2}\gamma + \frac{1+\sqrt{3}i}{3\cdot2^{2/3}}\frac{\Delta_1}{\big(\Delta_2+\sqrt{\Delta_2^2+4\Delta_1^3}\Big)^{1/3}} - \frac{1-\sqrt{3}i}{6\cdot2^{1/3}}\big( \Delta_2 + \sqrt{\Delta_2^2+4\Delta_1^3}\big)^{1/3} \label{eq:lambda-2-exact}\\
    \nonumber \\
    \lambda_3 & = -\frac{m_1+m_2}{3m_1m_2}\gamma + \frac{1-\sqrt{3}i}{3\cdot2^{2/3}}\frac{\Delta_1}{\big(\Delta_2+\sqrt{\Delta_2^2+4\Delta_1^3}\Big)^{1/3}} - \frac{1+\sqrt{3}i}{6\cdot2^{1/3}}\big( \Delta_2 + \sqrt{\Delta_2^2+4\Delta_1^3}\big)^{1/3} \label{eq:lambda-3-exact}
\end{align}
\text{with}
\begin{align}
    \Delta_1 & = 3\Big( \frac{k}{m_1} + \frac{k}{m_2} + \frac{\gamma^2}{m_1m_2} \Big) - \frac{(m_1+m_2)^2}{m_1^2m_2^2}\gamma^2 \\
    \nonumber \\
    \Delta_2 & = 9k\gamma \frac{m_1^2-4m_1m_2+m_2^2}{m_1^2m_2^2} + \frac{-2m_1^3+3m_1^2m_2+3m_1m_2^2-2m_2^3}{m_1^3m_2^3}\gamma^3  \quad .
\end{align}

Recall that $t_{p_i} = m_i/\gamma$ and $\tau = 2\pi\sqrt{m_1 m_2/k(m_1+m_2)}$. In regimes 1 and 2, we have $t_{p_1} \approx t_{p_2} \gg \tau$, which implies:
\begin{align}
    \label{eq:Separation_of_TimeScales}
    \frac{m_1}{\gamma} \gg \sqrt{\frac{m_1m_2}{k(m_1+m_2)}} \quad &, \quad  \frac{m_2}{\gamma} \gg \sqrt{\frac{m_1m_2}{k(m_1+m_2)}} \\
    \nonumber\\
    1 \gg \frac{\gamma}{\sqrt{k}}\sqrt{\frac{m_2}{m_1(m_1+m_2)}} \quad &, \quad  1 \gg \frac{\gamma}{\sqrt{k}}\sqrt{\frac{m_1}{m_2(m_1+m_2)}}   \quad .
\end{align}
Since $t_{p_1} \approx t_{p_2}$, it follows that $m_1 \approx m_2$, thus we have:
\begin{align}
    1 \gg \sqrt{\frac{\gamma^2 m_2}{k m_1(m_1+m_2)}} \approx \sqrt{\frac{\gamma^2 m_1}{k m_2(m_1+m_2)}}\approx \sqrt{\frac{\gamma^2}{k(m_1+m_2)}} \approx \sqrt{\frac{\gamma^2}{k(2m_1)}} \approx \sqrt{\frac{\gamma^2}{k(2m_2)}}  \quad .\label{eq:time-separation-regime1&2}
\end{align}

With Eq.~\ref{eq:time-separation-regime1&2}, we approximate $\lambda_i$ from Eqs.~\ref{eq:lambda-1-exact} - \ref{eq:lambda-3-exact}, keeping terms up to $O(k^{-1}m_j^{-2}\gamma^3)$ $[j \in 1,2 ]$, obtaining
\begin{align} 
\label{eq:Eignevalues_1_Approx}
    \lambda_1 & \approx \frac{-2\gamma}{m_1+m_2} - \frac{2(m_1-m_2)^2}{k(m_1+m_2)^4}\gamma^3\\
    \nonumber \\
    {\rm Re}(\lambda_2) = {\rm Re}(\lambda_3) & \approx \frac{-(m_1^2+m_2^2)}{2m_1m_2(m_1+m_2)}\gamma + \frac{(m_1-m_2)^2}{k(m_1+m_2)^4}\gamma^3 \\
    \nonumber \\
    {\rm Im}(\lambda_2) = -{\rm Im}(\lambda_3) & \approx \sqrt{\frac{k(m_1+m_2)}{m_1m_2}} - \sqrt{\frac{m_1m_2}{k(m_1+m_2)}}\frac{m_1^4+4m_1^3m_2-6m_1^2m_2^2+4m_1m_2^3+m_2^4}{8m_1^2m_2^2(m_1+m_2)^2}\gamma^2  \quad .\label{eq:Eignevalues_3_Approx}
\end{align}
Substituting Eqs. ~\ref{eq:Eignevalues_1_Approx} - \ref{eq:Eignevalues_3_Approx} into Eq.~\ref{eq:sigma_Diff} and keeping terms to $O(k^{-1}m_j^{0}\gamma^0)$, we have:
\begin{align}
    \sigma_{\Delta q_1}^2(\Delta t) & \approx \frac{1}{\beta} \Big[\Big( \frac{\Delta t}{\gamma} + \frac{1}{2k} - \frac{m_1+m_2}{2 \gamma^2}\Big) + \Big(\frac{m_1+m_2}{2\gamma^2} - \frac{m_1^2+2 m_1 m_2 - 3 m_2^2}{2 k (m_1+m_2)^2}\Big) \exp\Big(\frac{-2\gamma \Delta t}{m_1+m_2}\Big) \nonumber \\
    \nonumber\\
    & - \frac{2 m_2^2}{k(m_1+m_2)^2}\cos\Big(\sqrt{\frac{k(m_1+m_2)}{m_1 m_2} \Delta t}\Big)\exp\Big(-\frac{(m_1^2 + m_2^2) \gamma \Delta t}{2 m_1 m_2 (m_1+m_2)} \Big) \Big]  \quad .\label{eq:approximate-sigma-diff-1}
\end{align}
Furthermore, in regimes 1 and 2 we have $t_{p_1} \approx t_{p_2} \gg \Delta t$, and thus $\gamma \Delta t/m_j \ll 1$. Therefore, expanding the exponentials in Eq.~\ref{eq:approximate-sigma-diff-1} to $O(\gamma^2\Delta t^2 m_j^{-2})$, we obtain
\begin{align}
\label{eq:sigma_Approx_Regime1&2_Diff}
    \sigma_{\Delta q_1}^2(\Delta t) & \approx \frac{1}{\beta}\Big( \frac{m_2 \tau^2}{2\pi^2 m_1(m_1+m_2)}\big(1-\cos(2\pi \Delta t/\tau) \big) + \frac{\Delta t^2}{m_1+m_2} \Big) \quad .
\end{align}
Finally, using Eq.~\ref{eq:mstar_Exact_Diff_appendix}, we arrive at an approximate expression for $m^*$ that is valid in regimes 1 and 2:
\begin{align}
\label{eq:mstar_Approx_Regime1&2_Diff}
    m^* & \approx  \frac{(m_1+m_2)2\pi^2 (\Delta t/\tau)^2}{2\pi^2 (\Delta t/\tau)^2 + (m_2/m_1)(1-\cos(2 \pi \Delta t/\tau))} \quad .
\end{align}
As a consistency check, we confirm that that if $m_1 = m_2 = m$, then Eq.~\ref{eq:mstar_Approx_Regime1&2_Diff} reduces to Eq.~\ref{eq:mstar_Approx_Regime1&2_Same}.

\end{document}